\documentclass[sigconf]{acmart}

\usepackage{multirow}
\usepackage{subfigure}
\usepackage{enumitem}
\usepackage{dsfont}
\usepackage{diagbox}
\usepackage{hyperref}

\newcommand{\footref}[1]{\textsuperscript{\ref{#1}}}

\AtBeginDocument{%
  \providecommand\BibTeX{{%
    \normalfont B\kern-0.5em{\scshape i\kern-0.25em b}\kern-0.8em\TeX}}}

\copyrightyear{2020} 
\acmYear{2020} 
\setcopyright{acmcopyright}\acmConference[KDD '20]{Proceedings of the 26th ACM SIGKDD Conference on Knowledge Discovery and Data Mining}{August 23--27, 2020}{Virtual Event, CA, USA}
\acmBooktitle{Proceedings of the 26th ACM SIGKDD Conference on Knowledge Discovery and Data Mining (KDD '20), August 23--27, 2020, Virtual Event, CA, USA}
\acmPrice{15.00}
\acmDOI{10.1145/3394486.3403137}
\acmISBN{978-1-4503-7998-4/20/08}

\begin{document}
\fancyhead{}

\title{Towards Automated Neural Interaction Discovery for Click-Through Rate Prediction}
\titlenote{A majority of this work was done while the first author was interning at Facebook.}

\author{Qingquan Song$^1$, Dehua Cheng$^2$, Hanning Zhou$^2$, Jiyan Yang$^2$, Yuandong Tian$^2$, Xia Hu$^1$}

\affiliation{
  \institution{$^1$Texas A\&M University, College Station, TX} 
  \institution{$^2$Facebook Inc. Menlo Park, CA}
}
\email{{song_3134, xiahu}@tamu.edu, {dehuacheng, hanningz, chocjy, yuandong}@fb.com}

\begin{abstract}
Click-Through Rate (CTR) prediction is one of the most important machine learning tasks in recommender systems, driving personalized experience for billions of consumers. Neural architecture search (NAS), as an emerging field, has demonstrated its capabilities in discovering powerful neural network architectures, which motivates us to explore its potential for CTR predictions. Due to 1) diverse unstructured feature interactions, 2) heterogeneous feature space, and 3) high data volume and intrinsic data randomness, it is challenging to construct, search, and compare different architectures effectively for recommendation models. To address these challenges, we propose an automated interaction architecture discovering framework for CTR prediction named AutoCTR. 
Via modularizing simple yet representative interactions as virtual building blocks and wiring them into a space of direct acyclic graphs, AutoCTR performs evolutionary architecture exploration with learning-to-rank guidance at the architecture level and achieves acceleration using low-fidelity model. Empirical analysis demonstrates the effectiveness of AutoCTR on different datasets comparing to human-crafted architectures. The discovered architecture also enjoys generalizability and transferability among different datasets.
\end{abstract}





\maketitle

\section{Introduction}\label{intro}

Predicting Click-Through Rate (CTR) is a crucial problem in many web applications such as real-time bidding, display advertising, and search engine optimization~\cite{mcmahan2013ad}. Due to the large-scale dataset and high-cardinality feature property, extensive efforts have been devoted to designing architectures for effectively learning combinatorial feature interactions towards condensed low-dimensional feature representations~\cite{guo2017deepfm, lian2018xdeepfm, cheng2016wide, wang2017deep}.

Classical approaches usually put the efforts on designing explicit feature interactions and compose it with implicit interactions learning from multi-layer perceptrons (MLP) into a two-tower model~\cite{guo2017deepfm, wang2017deep, song2019autoint}. Beyond simple stacking strategies, how to organically bond the explicit and implicit interactions is still underexplored and may further promote effectiveness. Besides, existing work has shown that diamond MLP structure may be more powerful compared to the triangle and rectangular MLP structures~\cite{zhang2016deep}, which motivates us to explore more powerful implicit interactions via designing delicate MLP structures. In addition, conjoining the advantages of diversified explicit feature interactions such as inner and outer products could potentially boost the performance owning to the ensemble effect.

Neural architecture search (NAS) has emerged as a prevailing research field upon the prevalent adoption of deep learning techniques. It aims to discover optimal deep learning solutions automatically given a data-driven problem, thereby enabling practitioners to access the off-the-shelf deep learning techniques without extensive experience, and alleviating data scientists from the burden of manual network design. The rapid development of NAS research and systems has enabled the automation of state-of-the-art deep learning tools for various learning tasks in computer vision (CV) and natural language processing~\cite{elsken2019neural, chen2019techniques}. This motivates us to explore its potential in the context of tabular data in discovering complex neural interactions, specifically for CTR predictions.

Developing novel NAS approaches for neural interaction discovery and better CTR models is technically challenging. First, different from structure image data in CV tasks, CTR features are often heterogeneous, high-dimensional, and have both sparse and dense components, which are structureless and of diversified meanings in reality. Second, different from the dominant convolutional neural networks in CV tasks that consist of multiple structured convolutional operations, existing models for CTR prediction usually adopt multiple diverse and ad-hoc operations, leading to unstructured search space. Third, a practical model for CTR prediction is often trained on billions of data (e.g., Facebook has millions of daily active users and over 1 million active advertisers~\cite{he2014practical}, yielding billions of instances),
requiring the NAS process to be time and space efficient. Finally, the performance of CTR models with different architectures are often quite close in practice~\cite{wang2017deep,song2019autoint}, asking for the NAS approach to be sensitive and discriminating.

To cope with these challenges, we propose an automated neural interaction discovering framework for CTR prediction named \emph{\bf AutoCTR}. We abstract and modularize simple yet representative operations in existing CTR prediction approaches to formulate a generalizable search space. A hybrid search algorithm, composed of an evolutionary backbone and a learnable guider, is designed to perform orientated exploration. To enhance the exploitation power and balance the trade-off among different search objectives, we utilize a learning-to-rank strategy among the architecture level to filter out the locally inferior architectures, and conduct the survivor selection based on a mixture of rank-based measurement including aging, accuracy as well as architecture complexity. The search speed is further accelerated through a composited strategy of low-fidelity estimation, including data subsampling and hash size reduction. The main contributions are summarized from the following aspects:
\begin{itemize}[leftmargin=*]
    \item Design \emph{virtual blocks} and a hierarchical search space for CTR prediction by abstracting and unifying the commonly used operations in the existing literature.
    \item Provide ranking consistency analysis for three strategies combining low-fidelity estimation and weight inheritance. Empirically prove the availability of employing them for search acceleration.
    \item Propose a novel multi-objective evolutionary search algorithm with architectural-level learning-to-rank guidance.
    \item Empirically demonstrate the effectiveness of AutoCTR on different datasets comparing to human-crafted architectures, and validate the generalizability and transferability of the discovered architecture across different datasets.
\end{itemize}

\section{Preliminaries}
\noindent {\bf Click-Through Rate Prediction:} The CTR prediction problem could be mathematically defined as: given a dataset $D$ = $\{{\bf X}, {\bf y} \}$, where ${\bf X}\in \mathbb{R}^{N \times d}$ denotes the d-dimensional features matrix of $N$ instances. The features here consist of both sparse and dense features, where we assume the sparse features are ordinally encoded as integer vectors. ${\bf y}\in \{0, 1\}^{N}$ indicates the clicks of users to items. The goal is to predict the probability of a user clicking a target item.

Since its inception, the mainstream models have roughly experienced three-stage evolution starting from linear regression and tree-based models~\cite{mcmahan2013ad,he2014practical, yan2014coupled}, to interaction-based models~\cite{rendle2010factorization,juan2016field}, and then the deep neural networks (DNNs)~\cite{liu2015convolutional, qu2016product, zhang2016deep, cheng2016wide, guo2017deepfm, wang2017deep, he2017neural, lian2018xdeepfm, zhou2018deep, naumov2019deep}. Recent DNN-based work usually combines DNN with interaction-based or tree-based models to extracting condensed representations via learning combinatorial feature interactions.

\noindent {\bf Neural Architecture Search:}
Neural architecture search (NAS) aims to promote the design of neural network architectures automatically. Designing the NAS algorithm requires the specification of three main components, including \textit{search space}, \textit{search techniques}, and \textit{performance estimation strategy}~\cite{elsken2019neural}.  

From the search space perspective, existing work could be roughly divided into the exhaustive-architecture search space~\cite{jin2019auto, pmlr-v70-real17a} and the constrained cell space~\cite{zoph2018learning, pmlr-v80-pham18a, real2019regularized, liu2018darts}. The former one provides a more diversified set of architectures and allows a more comprehensive exploration, while the later one inductively limits the space to accelerate the search speed and reduce the search variance. Some tailored spaces are also designed~\cite{autodeeplab} for specific tasks. 

From the perspective of search techniques, several dominant ones include Bayesian optimization~\cite{jin2019auto}, reinforcement learning~\cite{zoph2017neural, zoph2018learning, pmlr-v80-pham18a}, evolutionary algorithms~\cite{pmlr-v70-real17a, real2019regularized}, gradient-based optimization~\cite{liu2018darts}, and tree-based methods~\cite{wang2019sample}. Recent work has shown the effectiveness of combining different types of search techniques to better balance exploitation and exploration~\cite{chen2019renas}.

Due to the high complexity of training the DNNs and the particularity of evaluation criteria in different tasks, various performance estimation strategies are proposed, including low-fidelity estimation~\cite{zoph2018learning,real2019regularized}, weight sharing~\cite{pmlr-v80-pham18a,liu2018darts}, learning curve extrapolation~\cite{wistuba2019inductive}, and network morphism~\cite{jin2019auto}.

\begin{figure}
\centering
\includegraphics[width=8.5cm]{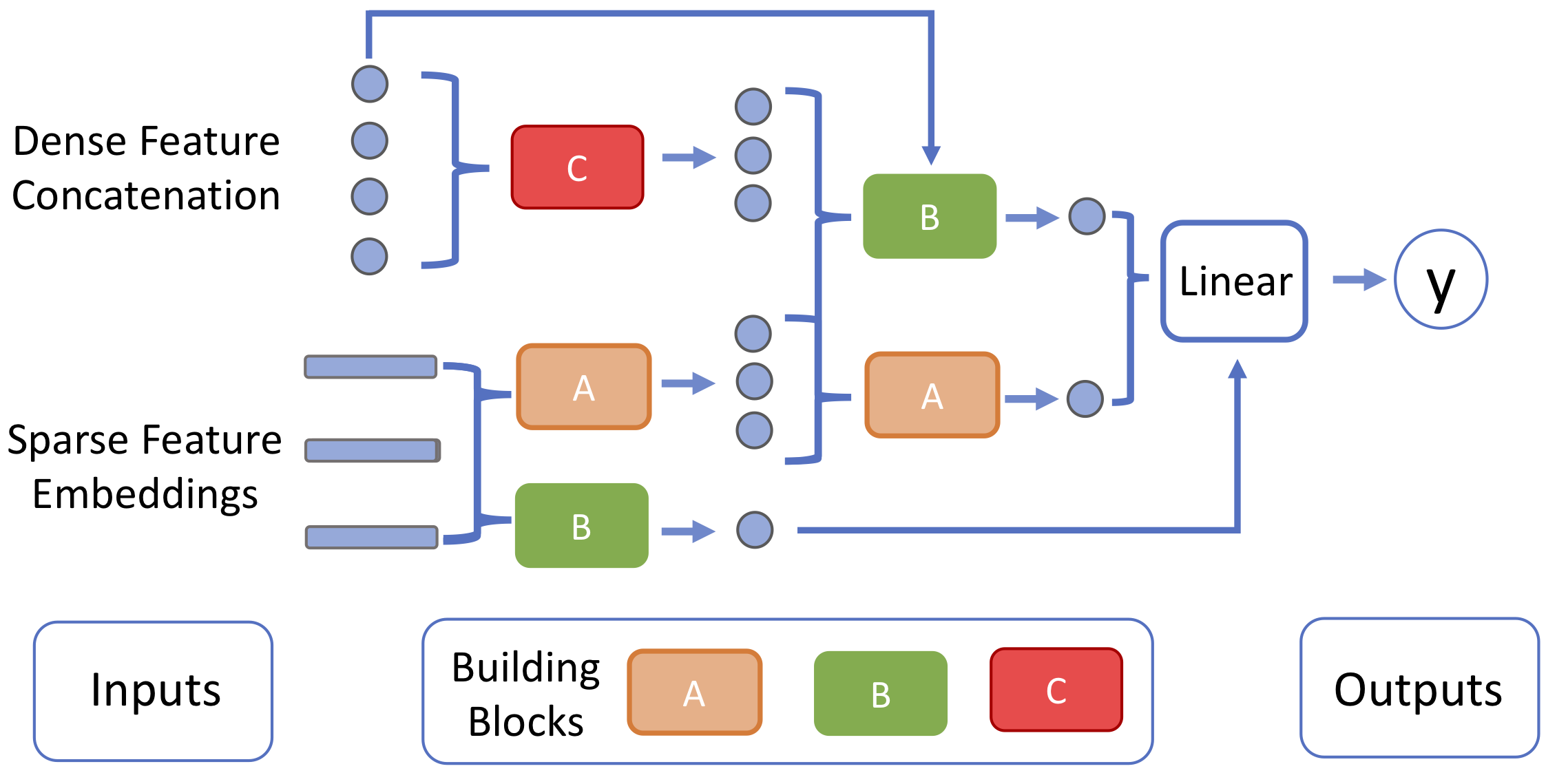} 
\caption{An illustration of an architecture in the designed search space. The virtual building blocks are wired together to form a DAG. Blocks are allowed to be selected repetitively.}
\label{fig:space}
\vspace{-0.4cm}
\end{figure}

\section{Hierarchical Search Space Design}\label{sec:ss}

An ideal search space should contain sufficient distinct architectures while potentially encompassing superior human-crafted structures~\cite{zoph2017neural, jin2019auto}. Inspired by the search space tailored to the vision tasks~\cite{autodeeplab}, we design a two-level hierarchical search space by extracting and abstracting representative structures in existing human-crafted CTR prediction architectures into virtual blocks, and wire them together as a set of direct acyclic graphs (DAGs) with dimensionality alignment among features. The inner space is composed of the appendant hyperparameters of blocks, such as the number of units and layers in a multi-layer perceptron block, while the connections among blocks form the outer search space.

As shown in Figure~\ref{fig:space}, for a given instance, we assume its raw input dense features are concatenated into a vector, and its sparse features are embedded into low-dimensional vectors based on the look-up-table operation following similar preprocessing done in various CTR models~\cite{wang2017deep, naumov2019deep}. Blocks are wired together to form a DAG, and each block could take both raw input features and the outputs from blocks with higher topological order via feature concatenation and dimensionality alignment. The final block of the network is set to be a linear transformation. It collects all the untouched features from either raw inputs or outputs provided by other blocks. It is worth noting that to simplify the setting, the following hyperparameters are not taken into account in the search space: (1) Hash size of sparse features. (2) Embedding dimension of sparse features. (3) Optimizers and other model training hyperparameters such as learning rate and batch size.

\subsection{Virtual Block Abstraction}

We select and extract building blocks by considering two aspects described as follows:
\begin{itemize}[leftmargin=*]
\item {\bf Functionality:} blocks should accommodate and complement each other. Each type of block should accommodate both dense and sparse features as inputs, and can be quantitatively evaluated.

\item {\bf Complexity Aware:} The computational and memory cost of a block should be a simple function of its input specification and hyperparameters. The involvement of these primitives could benefit the design of complexity-aware search algorithms to achieve better resource management and low-complex architectures. 
\end{itemize}

Upon these requirements, we could abstract various operations from existing work as blocks with different functionality and levels of complexity such as multi-layer perceptron (MLP)~\cite{naumov2019deep}, dot product (DP), factorization machine (FM)~\cite{rendle2010factorization}, outer product~\cite{wang2017deep}, and self-attention~\cite{song2019autoint}, etc. We elaborate on the construction of the three example blocks (i.e., MLP, DP, FM) adopted in the experiments in Appendix~\ref{sec:block} and describe the way of aligning the dimension and accommodating different inputs for each of them. Other blocks could also be easily abstracted and integrated into the framework, which is left for future exploration.

\subsection{Summarization of Search Components}\label{sec:space}
We summarize the main components to be searched in the hierarchical search space as follows:

\begin{itemize}[leftmargin=*]
\item {\bf Block Type:} MLP, FM, DP.
\item {\bf Raw Feature Input Selection:} each block is allowed to take the raw feature with four choices, i.e., dense only, sparse only, both or none. Without particular emphasis, we group the raw input dense features as one single component to be selected rather than considering them independently. A similar procedure is done for the sparse features to reduce the search complexity.

\item {\bf Inter-Block Connection:} a block could receive the outputs from any block that appeared before it. The order is defined as the topological order in the DAG.
\item {\bf Block Appendant Hyperparameters:} We only consider the number of hidden units of MLP block in this work. The embedding sizes for aligning the input dimensions in different blocks are fixed and set to be the same with the embedding size of the raw input sparse features to reduce the number of parameters. 

\end{itemize}

To enable the feasible adoption of different searching algorithms, we provide a vector representation of each architecture as a concatenation of multiple block vectors following~\cite{wang2019sample}. Each block is vectorized as a concatenation of the four components, i.e., \textit{[Block Type, Raw Feature Input Selection, Inter-Block connection, Block Appendant Hyperparameters]}, where the \textit{Block Type} and \textit{Raw Feature Input Selection} are encoded as one-hot vectors respectively, \textit{Inter-Block connection} is encoded as a multi-hot vector, and \textit{Block Appendant Hyperparameters} is encoded as a vector of ordinals.

The designed search space contains plentiful distinct architectures. Even with three types of blocks to be selected and assume each architecture contains no more than seven blocks, the space would still contain over $10^{11}$ distinct architectures. Moreover, it could cover multiple representative human-crafted architectures such as deepFM~\cite{guo2017deepfm}, DLRM~\cite{naumov2019deep}, IPNN~\cite{qu2016product}, and Wide\&Deep~\cite{cheng2016wide}.

\section{Multi-Objective Evolutionary Search with Hyperrank Guidance}

The proposed searcher is a mixed searcher composed of an evolutionary algorithm and a learning-to-rank guider. We select the evolutionary algorithm in this pilot study due to its simplicity and effectiveness in balancing the exploitation and exploration~\cite{real2019regularized}. We adopt a multi-objective evolutionary searcher as the backbone and employ a tree-based learner to guide the mutation of a selected parent architecture in each iteration to facilitate the exploitation of superior offsprings. The search process could be described as a loop of three stages, i.e., parent selection, guided mutation, and survivor selection. The initial population is constructed via randomly selecting and evaluating a predefined number of architectures. Stratified selections could also be used to potentially enhance the performance, which we leave for future exploration.

\begin{figure}
\centering
\includegraphics[width=3.35in]{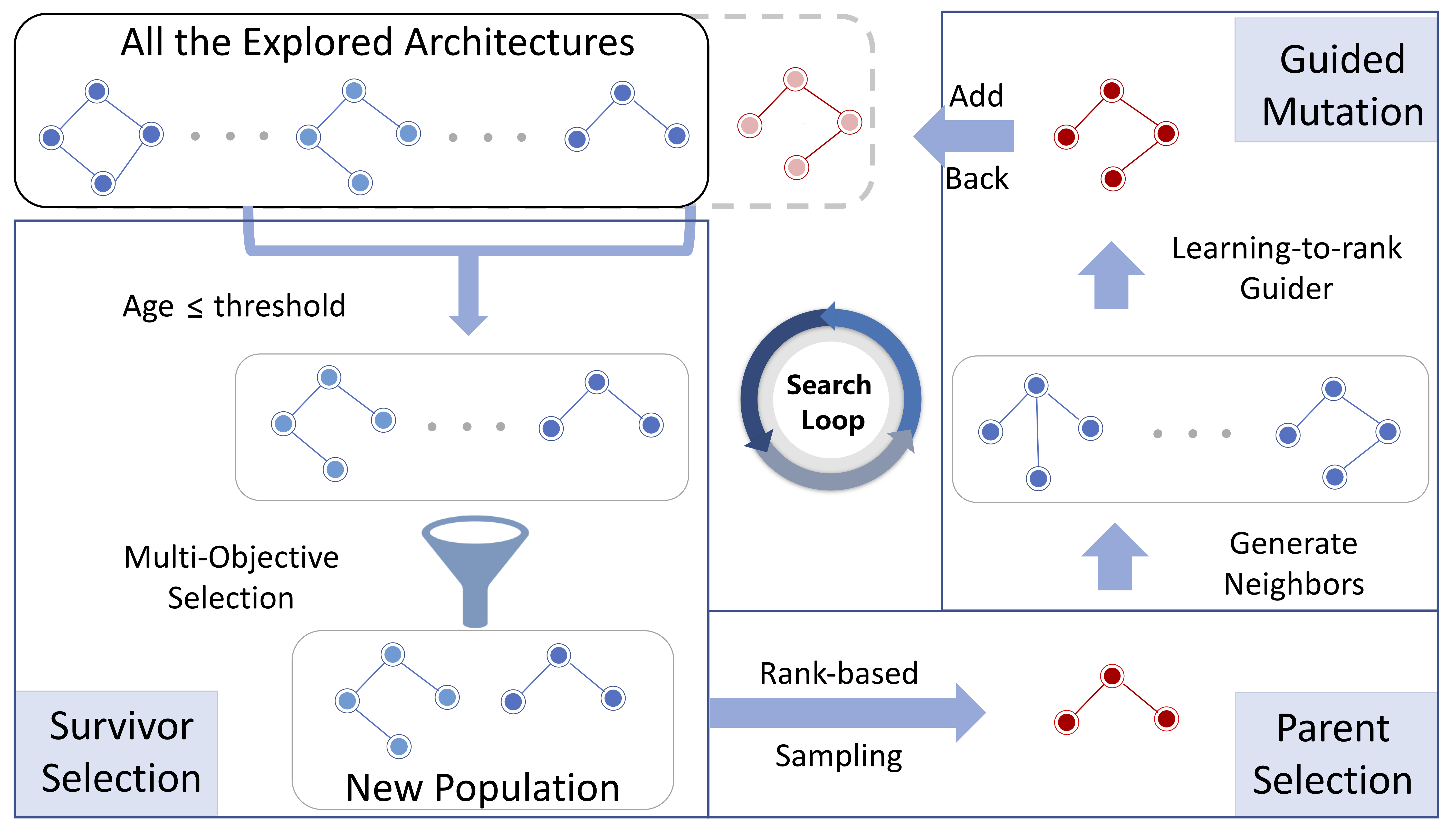} 
\vspace{-0.6cm}
\caption{An illustration of the AutoCTR search loop}
\label{fig:searcher}
\vspace{-0.3cm}
\end{figure}

The basic idea of the search loop is shown in Figure~\ref{fig:searcher}. In each search loop, we leverage a mixture of rank-based meta-features to perform the survivor selection and maintain a new fix-size population. Then we select a parent architecture from the population based on designed discrete probabilistic distribution. After that, a set of neighbors is generated via mutating the selected parent architecture. We adopt a learning-to-rank mechanism upon the architecture level and select the best offspring from the generated neighbors. After evaluating the performance of this offspring, we add it back to the architecture pool explored so far. The three stages are elaborated in turn in the following subsections. 

\subsection{Multi-Objective Survivor Selection}

To maintain a superior population with diversified architectures, we design a survivor selection metric $f$ to measure the survival value of each architecture and select the top-$p$ ones as the population for parent selection. Three types of objectives are considered in the metric, i.e., fitness, age, and model complexity. The ``fitness'' here represents the performance of an explored architecture to ensure the exploitation ability, while the ``age'' reflects the reverse order of the architectures explored so far. We use the rank of logloss as the fitness measure to mitigate the difference in scale and define the ``age'' as the existing time of an architecture explored so far, motivated from the age-based evolutionary methods~\cite{real2019regularized}, to enhance the exploration of diversified architectures. At each search loop, we set the current time to be 0 and set the age of each observed architecture as the number of architectures explored after it. In particular, all initial architectures are assigned with the same age. Besides the two objectives, since CTR tasks are usually resource hungry in practice, we also take the ``model complexity'' into account to explicitly constrain the model complexity during search. We adopt floating point operations per second (FLOPs) as the complexity metric and use the rank of flops to mitigate the scale influence. 

The designed selection schema consists of two steps as shown in Figure~\ref{fig:searcher}, which could be formulated as follows:
\begin{equation}\label{equ:multiobj}
    f(q, a_A, r_A^q, c_A^q) = \mathds{1}_{[a_A\le q]} \cdot (\mu_1 a_A + \mu_2 r_A^q + \mu_3 c_A^q),
\end{equation}
where $a_A$ denotes the age of an architecture $A$. The indicator function   $\mathds{1}_{[a_A\le q]}$ is used to filter out the architectures that is older than $q$, where $q$ is a hyperparameter larger than the population size $p$, such that the architectures with high-performance or low-complexity in the pool would not be selected consistently. $r_A^q, c_A^q \in \{ 1,2,\ldots q \}$ are the performance and complexity ranking of architecture $A$ within the $q$ ``youngest'' architectures, respectively. $\{\mu_i\}_{i=1,2,3}$ are the trade-off hyperparameters to balance the different objectives.

\subsection{Rank-Based Parent Selection}\label{sec:parent}
Suppose we maintain a population of size $p \in \mathbb{Z}^+$ after the survivor selection step. The goal of parent selection is to select an architecture from the population for generating a premium offspring to be evaluated. Several popular strategies used in conventional evolutionary algorithms include proportional, ranking, tournament, and genitor selection ~\cite{zhang2000comparison}. In this work, we adopt a ranking selection schema and design a nonlinear ranking distribution borrowing the idea from tournament selection. The intuition comes from three aspects: (1) The performance (logloss or AUC) of different architectures in CTR prediction tasks is often extremely close, and the scale may vary a lot on different datasets. It is hard to design a unified performance-based distribution to achieve adaptive selective pressure on different datasets. (2) Existing work has proved the effectiveness of ranking and tournament selection comparing with proportional and genitor on balancing the selective intensity and selection diversity~\cite{blickle1996comparison,zhang2000comparison}. (3) Classical tournament methods usually randomly select a fixed number of candidates first and then select the best one of them. This may result in a portion of architectures in the population never being selected as the parent, especially when the ratio between candidate size and population size is large.

The concrete design of the probability is as follows:
\begin{equation}
    p(r^\ast_A) = \frac{ {r^\ast_A + \lambda - 1 \choose \lambda} }{ {p + \lambda \choose  1 + \lambda }}, \qquad r^\ast_A \in \{1, 2, \ldots, p\}, \lambda \in \mathbb{N}^0,
\end{equation}
where $r_A^\ast$ denotes the rank of an architecture $A$ in the population, ${n \choose k} = \frac{n!}{(n-k)!k!}$, $(k=0,1,.., n)$ , and $\lambda$ is a hyperparameter to balance the trade-off between selection intensity and selection diversity. Given a fixed size of population, with the increasing of $\lambda$, the selection intensity increases while the selection diversity reduces. In particular,  $\lambda=0$ is the same as uniform selection and when $\lambda=1$ is the same as linear rank selection~\cite{back1994selective, blickle1996comparison}.

\subsection{Guided Mutation by Learning to Hyperrank}

After the parent architecture is selected, the last step is to generate a worth exploring offspring upon it. A naive way of doing this is to randomly select and modify an operation in the parent architecture~\cite{real2019regularized}. Nevertheless, it could become an inefficient strategy due to the huge search space and the waste of the architecture information explored so far. Existing work has demonstrated the effectiveness of using a learning-based model to guide the mutation process~\cite{chen2019renas}. However, with a limited number of explored architectures and the extremely close performance among them, it is non-trivial to learn an effective fitness-based guider.

\begin{figure*}[t]
\centering
\vspace{-0.2cm}
\hspace{5pt}
\subfigure[Rank consistency varying with the number of subsamples and different training strategy]{
\includegraphics[width=.3\textwidth]{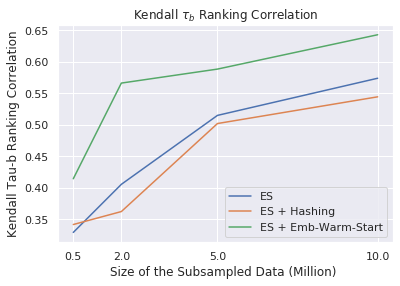}
    \label{fig:rank1}
}
\hspace{5pt}
\subfigure[Local ranking consistency shifting with window size 30 upon the sorted ``ground-truth'' rank]{
\includegraphics[width=.3\textwidth]{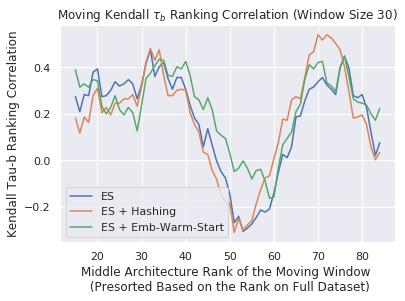}
    \label{fig:rank2}
}
\hspace{5pt}
\subfigure[NDCG@K score of different training strategies on 2 million subsamples]{
\includegraphics[width=
.3\textwidth]{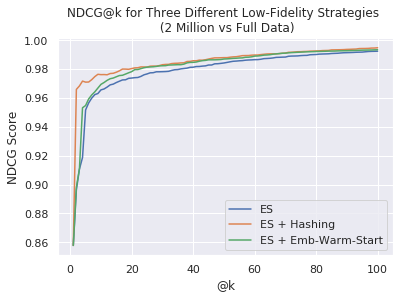}
    \label{fig:rank3}
}
\vspace{-0.2cm}
\caption{Rank consistency analysis from both global perspective and local perspective}
\vspace{-0.2cm}
\label{fig:para} 
\end{figure*}

Instead of learning a fitness-based guider, we adopt a learning-to-rank strategy to learn the relative ranking among architectures based on a pairwise ranking loss and the gradient boosted tree learner. The whole offspring generation process is done via three steps:  (1) train a guider based on the exploitable dataset (e.g., all the architectures explored so far); (2) randomly generate a set of unique neighbors around the parent architecture; (3) select the best neighbor as the offspring based on the guider. We call it ``learning-to-Hyperrank'' as it conducts a model-level ranking. The intuition comes from two perspectives. On the one hand, the search problem itself is essentially a ranking problem. Learning the ranking relationship is a commensurable strategy comparing with learning the fitness but is weaker and more flexible.  On the other hand, the number of instances are implicitly augmented from the point-wise inputs to the pairwise inputs.

For the pairwise ranking-loss, we use LambdaRank~\cite{burges2010ranknet} due to its simplicity and efficiency. Though different types of models could serve as the learner, we choose a gradient boosted tree learner as an example here due to its general stable and superior performance on small-scale datasets. To feed architectures as the input for the tree-based learner, we encode each block as a vector and concatenate them based on their topological order. Each block vector follows the vector representation described in section~\ref{sec:space}.

\section{Performance Estimation Acceleration}\label{sec:lf}

One of the most crucial challenges in modern NAS research, which could be even more severe in recommender systems, is the high time complexity of network training. Low-fidelity performance estimation and weight inheritance are two of the most widely adopted methods for speeding up performance estimation~\cite{zoph2018learning,real2019regularized}. However, extra bias could be introduced, leading to the variation of relative ranking among architectures, thereby affecting the searching effectiveness~\cite{elsken2019neural}. We consider two strategies of low-fidelity estimation and adopt a warm-start embedding trick leveraging the weight inheritance among architectures to mitigate the time and resource complexity. These strategies are all general and practical ways to speed up the manual tuning of recommender systems.  Several rank consistency tests are described afterward to provide the evidence and illustrate the feasibility of adopting these methods.

\begin{itemize}[leftmargin=*]
\item {\bf Data Subsampling.} For each dataset, we randomly subsample a predefined portion of data for searching and transfer the searched best architecture on the full dataset for final evaluations. On average, the training time of every single architecture in our experiment is roughly linearly correlated with the subsampling ratio under the hardware setting described in Appendix~\ref{sec:hardware}.

\item {\bf Reducing Hash Size.} For high-level categorical features ($>10^4$), we hash the cardinality of them to $10^4$ before the embedding step to reduce the embedding size, and set back to the original cardinality during final fit on the full dataset.

\item {\bf Warm-Start Embedding.} We pretrain a simple three-layer MLP model (units: 128-1024-128) on the full dataset and use the pretrained embeddings of sparse features as the warm-start for each architecture before the low-fidelity training.

\end{itemize}

The rest of this section provides an analysis of the effect of adopting low-fidelity training on ranking consistency. We use logloss as the evaluation metric and use the early-stopping (ES) strategy to alleviate overfitting. We focus on the global and local rank consistency respectively, to pursue the analysis. The global rank consistency inspects whether the estimation could reflect the actual ranking of performance among architectures\footnote{We assume the actual rank of architectures is reflected by their high-fidelity performance achieved on the full dataset.}, and the local rank consistency testing zooms into the architectures with relatively closer performance and analyze their localized rank consistency.

\subsection{Global Rank Consistency}

{\bf Settings:} We use Criteo dataset\footnote{\label{criteo}\url{www.kaggle.com/c/criteo-display-ad-challenge}} here for experiments. Without loss of generality, we narrow down the search space by assuming each architecture contains five blocks, and each MLP block has one layer with $128$ units. We randomly sample 100 valid architectures and evaluate them on Criteo with different sizes of subsamples, i.e., $\{0.5, 2, 5, 10\}$ million. Each subsampled dataset is split into training ($80\%$), validation ($10\%$), and test ($10\%$) sets. We run the $100$ models on the full dataset three times to provide a ``ground-truth'' rank, and the rank consistency is measured by the Kendall $\tau_b$ coefficient ranging from $-1$ (perfect inversion) to $1$ (perfect agreement).

\noindent{\bf Observations:} Figure~\ref{fig:rank1} depicts the varying curves of three training strategies measured by the Kendall $\tau_b$ coefficient with the increasing of the subsample size. The three strategies are: (1) early stopping; (2) early stopping with sparse feature hashing; (3) early stopping with warm-start embedding. We observe that: firstly, with the increasing of the subsample size, the rank among architectures become more consistent with the “ground-truth” rank, and the growth speed gradually becomes slow.  The non-linear relationship  between sample size and $\tau_b$ coefficient offers the opportunity of adopting the low-fidelity setting during searching, which will be further evaluated in the experimental section. Secondly, adopting the hashing strategy for sparse features with high cardinality would marginally decrease the rank consistency. Finally, the warm-start embedding strategy could increase ranking consistency.

\begin{figure*}[t]
\centering
\includegraphics[width=7in]{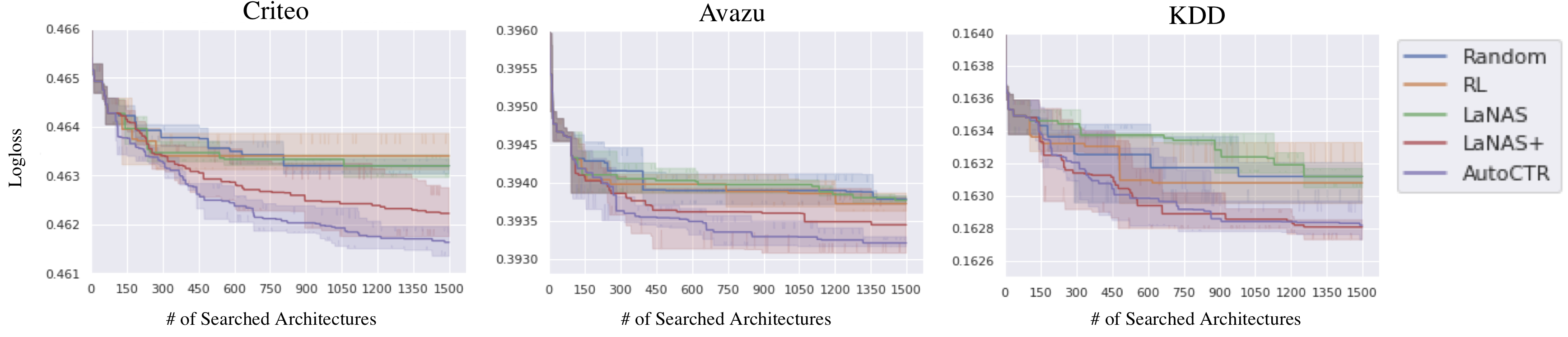} 
\vspace{-0.8cm}
\caption{ The performance drifting of the best architecture during the search process on different datasets}
\label{fig:general}
\end{figure*}

\subsection{Local Rank Consistency}
The above analysis provides a macro view of the ranking consistency among the architectures. Beyond this, we are also curious if the $\tau_b$ score is harmoniously distributed across different intervals of the actual rank. We first sort the 100 architectures based on their “ground-truth” performance evaluated on the full dataset. Better architectures are indexed with smaller numbers. Then we depict two plots for analysis. Figure~\ref{fig:rank2} displays the variation of $\tau_b$ coefficient with a length-$30$ sliding window among the sorted architectures.  For example, the score of the first point denotes the $\tau_b$ coefficient of the top-30 architectures, where its x-coordinate is $15$, indicating the middle architecture rank within this sliding window. Figure~\ref{fig:rank3} shows the variation of the NDCG@k score with the increase of $k$. Exponential gain is used to calculate the NDCG score.

From the two figures, two main observations could be found: (1) From Figure~\ref{fig:rank2}, we can see that, best- and worst-performing architectures seem to be more locally rank consistent. This is partially aligned with our expectations since we expect the top architectures and the bottom ones to be more easily discernible than others. The warm-start strategy is more helpful for maintaining the local rank of middle architectures rather than the polar ones. (2) From Figure~\ref{fig:rank3}, we can observe that the rank of the top architectures could generally maintain high rankings in the low-fidelity setting even if the rank consistency for some mediocre architectures is affected more compared to the polar ones.

\section{Experimental Analysis}\label{sec:exp}

In this section, we empirically evaluate AutoCTR as well as several baseline searchers and compare the discovered architecture to the human-crafted architectures. We use logloss and AUC score as the core evaluation metrics. Four questions are mainly explored:

\begin{enumerate}[leftmargin=.25in]

    \item[{\bf Q1.}] How is AutoCTR comparing with other baseline searchers on both the search efficiency and effectiveness?

    \item[{\bf Q2.}] How is the performance of the best architecture explored by the AutoCTR comparing with the state-of-the-art (SOTA) human-crafted architectures?

    \item[{\bf Q3.}] Are the searched architectures able to be transferred between different datasets?

    \item[{\bf Q4.}] How sensitive is the AutoCTR to its key hyperparameters?

\end{enumerate}

\subsection{Baselines}
We select baselines from both NAS methods and human-crafted CTR architectures for comparison. Since several searchers are not directly applicable in our setting, we modify and improve their flexible components and elaborate the details in the Appendix~\ref{sec:baseline}. The selected NAS methods include random search, reinforcement learning based search (RL), and a sample-efficient tree-based search: latent action neural architecture search (LaNAS)~\cite{wang2019sample}. We also include a variation of LaNAS named LaNAS+, which is improved from two perspectives. (1) We borrow the idea from AutoCTR to perform evolutionary sampling in LaNAS after a narrowed search space is selected, which is originally done by random. It highly improves the search speed and the exploitation power upon the experimental results. (2) As LaNAS requires each architecture to be finished in order to explore a new one, we include virtual loss to enable its parallelization.  See Appendix~\ref{sec:baseline} for more details. Some other searchers are not directly considered due to the complexity of converting them to fit our search space, such as the kernel design and adjustment in the Bayesian optimization approach (e.g.,~\cite{jin2019auto}), and the super graph construction in gradient-based methods (e.g.,~\cite{liu2018darts}), which are left for future exploration. We select three representative human-crafted networks DLRM~\cite{naumov2019deep}, DeepFM~\cite{guo2017deepfm}, and AutoInt+~\cite{song2019autoint}, in which the first two are covered in our search space. More details about the baselines are elaborated in Appendix~\ref{sec:baseline}.

\subsection{Experimental Settings}

\subsubsection{Data Preprocessing.} We adopt three benchmark datasets in this paper, i.e., Criteo\footref{criteo}, Avazu\footnote{\label{avazu}\url{https://www.kaggle.com/c/avazu-ctr-prediction/data}}, KDD Cup\footnote{\label{kdd}\url{https://www.kaggle.com/c/kddcup2012-track2/data}}. The basic statistics of them are summarized in Table~\ref{tab:dataset} in Appendix~\ref{sec:data}. All three datasets are processed based on the way and codes provided in~\cite{song2019autoint}. During the search phase, we subsample the first $2$ million data of each dataset and further divide it into the tiny training (80\%), validation (10\%), and test (10\%) sets for low-fidelity evaluation.

\subsubsection{Hyperparameter Settings.} We search seven intermediate blocks for each architecture. Blocks are allowed to be empty. Three example blocks are adopted in the final experiments, i.e., MLP, FM, and DP. The detailed construction of them is provided in Appendix~\ref{sec:block}. We randomly sample 100 architectures as initialization for all the searchers. For the RL searcher, we adopt the off-policy training to go through the 100 architectures. For AutoCTR, the 100 architectures will directly form the initial population. For LaNAS and LaNAS+, the initial tree splits are learned with these 100 architectures. We repeat the experiments three times with three different seeds. Detailed hyperparameter settings for each searcher and the network training in both the search phase and final fit phase, are specified in Appendix~\ref{sec:expset} due to the page limitation.

\begin{table*}[t]\scriptsize
\vspace{-0.1cm}
\caption{General CTR prediction results on the three benchmark datasets}
\vspace{-0.4cm}
\label{table:performance}
\begin{center}
\begin{tabular}{c|c|c c c c c c|c}
\toprule
& &  \multicolumn{2}{c}{Criteo} & \multicolumn{2}{c}{Avazu} &  \multicolumn{2}{c}{KDD} & \multirow{2}{*}{\shortstack[c]{Search cost \\ (GPU Days)}}
\\  %
&  &  Logloss & AUC  & Logloss & AUC   & Logloss & AUC \\\hline
\multirow{3}{*}{\bf \shortstack[c]{SOFA \\ human-crafted \\ Networks}}     
					 &  DeepFM    & 0.4432  &  0.8086 & 0.3816  & 0.7767  & 0.1529 & 0.7974 & -  \\  
					 &  DLRM      & 0.4436  &  0.8085 & 0.3814 & 0.7766 & 0.1523 & 0.8004 & - \\
					 &  AutoInt+  & 0.4427 &  0.8090  & 0.3813 & 0.7772  & 0.1523  & 0.8002 & - \\
					 \midrule
\multirow{6}{*}{\bf \shortstack[c]{Best Networks  \\ Found by the \\ NAS Methods}}     
                    &  Random     & 0.4421 {\footnotesize $\pm$ 0.0003} &  0.8096 {\footnotesize $\pm$ 0.0004} & 0.3824 {\footnotesize $\pm$ 0.0030} & 0.7765 {\footnotesize $\pm$ 0.0029} & 0.1531 {\footnotesize $\pm$ 0.0001} & 0.8001 {\footnotesize $\pm$ 0.0003} & $\sim$ 0.75 \\ 
				    &      RL     & 0.4422 {\footnotesize $\pm$ 0.0005} &  0.8094 {\footnotesize $\pm$ 0.0005} & 0.3810 {\footnotesize $\pm$ 0.0003} & 0.7778 {\footnotesize $\pm$ 0.0005} & 0.1531 {\footnotesize $\pm$ 0.0001} & 0.7999  {\footnotesize $\pm$ 0.0002} & $\sim$ 0.75 \\ 
					&    LaNAS    & 0.4421 {\footnotesize $\pm$ 0.0004} &  0.8096 {\footnotesize $\pm$ 0.0005} & 0.3814  {\footnotesize $\pm$ 0.0006} & 0.7772 {\footnotesize $\pm$ 0.0011} & 0.1533 {\footnotesize $\pm$ 0.0002} &  0.8001 {\footnotesize $\pm$ 0.0009} & $\sim$ 5 \\ 
					&    LaNAS+   & 0.4417 {\footnotesize $\pm$ 0.0001} &  0.8101 {\footnotesize $\pm$ 0.0000} & 0.3800 {\footnotesize $\pm$ 0.0004} & 0.7790 {\footnotesize $\pm$ 0.0007} & 0.1521 {\footnotesize $\pm$ 0.0001} & 0.8009 {\footnotesize $\pm$ 0.0004} & $\sim$ 0.75 \\ 
					&    AutoCTR  & \textbf{0.4413 {\footnotesize $\pm$ 0.0002} }  &  \textbf{0.8104  {\footnotesize $\pm$ 0.0003} }  & \textbf{0.3800 {\footnotesize $\pm$ 0.0001}} & \textbf{0.7791 {\footnotesize $\pm$ 0.0001} } & \textbf{0.1520 {\footnotesize $\pm$ 0.0000}} & \textbf{0.8011 {\footnotesize $\pm$ 0.0001}} & $\sim$ 0.75 \\ 
					
					&    AutoCTR (warm)  & 0.4417 {\footnotesize $\pm$ 0.0005}  &  0.8099  {\footnotesize $\pm$ 0.0005}  & 0.3804 {\footnotesize $\pm$ 0.0004} & 0.7784 {\footnotesize $\pm$ 0.0006} & 0.1523 {\footnotesize $\pm$ 0.0001} & 0.8004 {\footnotesize $\pm$ 0.0003} & $\sim$ 0.75 \\ 
\bottomrule
\end{tabular}
\end{center}
\vspace{-0.3cm}
\end{table*}

\begin{table}[t]\small
\caption{Architecture complexity comparison (parameters in the embedding tables are included)}
\label{table:complexity}
\begin{center}
\vspace{-0.18cm}
\begin{tabular}{c|ccc ccc}
\toprule
 & \multicolumn{3}{c}{\# Params (Million)} & \multicolumn{3}{c}{Flops (Million)} \\
    &  Criteo   & Avazu & KDD    &  Criteo   & Avazu & KDD  \\\hline
					 DeepFM     &  22.51 & 30.34 & 101.73    & 22.74 & 22.50 & 21.66 \\  
					 DLRM      &  23.55 & 29.29 & 102.77    & 26.92 & 18.29 & 25.84 \\
					  AutoInt+   &  20.44 & 28.28 & 99.66     & 18.33 & 17.49 & 14.88 \\
					 \midrule
    
					 AutoCTR  & 19.89 & 26.49 & 97.06 &  12.31 & 7.12 & 3.02 \\ 
\bottomrule
\end{tabular}
\end{center}
\vspace{-0.4cm}
\end{table}

\subsection{General Comparison Among Searchers}

We first compare the general search performance of all five searchers. Figure~\ref{fig:general} depicts the logloss drifting of the best architecture during searching on the three datasets. The x-axis indicates the number of architectures searched so far (in total, 1500). The y-axis denotes the validation logloss of the architectures. Several observations can be summarized as follows. Firstly, based on the performance of the best architecture searched in the low-fidelity setting, AutoCTR generally outperforms other baselines, and our modified LaNAS+ outperforms LaNAS. Secondly, by comparing the search efficiency, AutoCTR and LaNAS+ still outperform other searchers consistently.

We then transfer the best architectures searched so far by each searcher onto the full dataset and display the final evaluation results in Table~\ref{table:performance}. The results shown from row four to eight are the average performance of the best architectures searched in the three rounds with different seeds. We can observe that: (1) After transferring the best architectures found by the searchers on the full datasets, the architectures found by AutoCTR still performs the best. Moreover, the ranking of the results aligns well with the one in low-fidelity setting, which implies the correctness of the rank consistency testing and the feasibility of adopting low-fidelity estimation in the search process. (2) On all three datasets, the final architectures searched by AutoCTR and LaNAS+ could achieve even better performance comparing to the SOTA architectures. The architectures searched by Random search and RL-based search could also achieve comparable or even better performance.  This empirically validates the effectiveness of the designed search space and the feasibility of adopting NAS algorithms on the CTR prediction problem. It is worth pointing out that an improvement of around $0.0005$-$0.001$ is already regarded as practically significant on these CTR prediction benchmarks~\cite{wang2017deep,song2019autoint}. (3) We further examine the influence of adopting the warm-start embedding in searching. Although it highly improves the performance of most architectures in the low-fidelity setting (not shown in Figure~\ref{fig:general} due to the difference in scale), the performance on the full dataset is not improved. We attribute this observation to the overfitting issue of the warm-start embedding dictionary and the evaluation dataset during searching. Since we set a fixed search iteration (i.e., 1500 architectures) rather than adopting early-stop for the search algorithm, the overfitting issue may happen during the search process, which has also been pointed out by several recent works~\cite{jiang2019neural, zela2020understanding}.

Beyond accuracy, we also display the time complexity of the search algorithms in Table~\ref{table:performance} and compare the model complexity of the best-discovered architectures of AutoCTR with and the SOTA human-crafted architectures in Table~\ref{table:complexity}. The time for training the searcher could generally be ignored due to the limited size of the sampled architectures and the parallel CPU-GPU training schema we adopted. Results show that the explored architecture is smaller than the human-crafted ones on both the number of parameters and FLOPs. This is mainly because: (1) we explicitly constrain search space and adopt the complexity control term in the survivor selection, which restricts the exploration of overcomplicated architectures; (2) the searchers tend to find architectures with diamond or inverted triangle MLP structures, while the human-crafted ones directly adopt rectangular MLP structures as defined in the original work, and are set with 1024 units in each layer in our experiments. 

\subsection{Architecture Transferability Analysis}

As the human-crafted architectures are not designed for a specific dataset, we explore the transferability of the searched architectures across different datasets. We select the best architectures searched by AutoCTR on each dataset, and apply them on the other two. From Table~\ref{tab:transfer}, we can observe that the architectures searched on one dataset could still perform well when applying to the others. This is mainly because: (1) the three benchmark datasets share common characteristics of feature relationships; (2) the blocks incorporated in the designed search space and the discovered connectivity among the blocks is general enough to uncover the high-order feature interactions of different CTR prediction datasets. Comparably, the architecture discovered on Avazu performs a bit worse when doing the transfer. One reason is that Avazu only contains sparse features, which results in no exploration of the dense features in searching. For these models, the dense features are only considered in the final embedding layer in our implementation during the transfer.

\begin{table}[t]\small
\caption{Transferability of architectures found by AutoCTR}
\vspace{-0.25cm}
\label{tab:transfer}
\begin{center}
\begin{tabular}{c|c|c|c}

\toprule
\backslashbox{\begin{tabular}[c]{@{}c@{}} Original Dataset \end{tabular}}{\begin{tabular}[c]{@{}c@{}} Target Dataset \end{tabular} }  & Criteo & Avazu & KDD \\\hline
Criteo  & 0.4413    & 0.3799   & 0.1520 \\\hline
Avazu   & 0.4421    & 0.3800   &    0.1535   \\\hline
KDD & 0.4418    & 0.3803   & 0.1521 \\\bottomrule
\end{tabular}
\end{center}
\vspace{-0.3cm}
\end{table}

\subsection{Hyperparameter Sensitivity Analysis}
\label{sec:hyper}

In this section, we study the sensitivity of AutoCTR on the key hyperparameters using the Criteo dataset and analyze the impact of the core components at different stages.

\begin{figure*}[t]
\centering
\vspace{-0.2cm}
\hspace{5pt}
\subfigure[Effects of the selection intensity hyperparameter $\lambda$ to the search process given population size $p$=100.]{
\includegraphics[width=.3\textwidth]{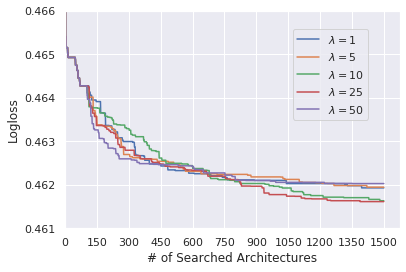}
    \label{fig:stage1}
}
\hspace{5pt}
\subfigure[Effects of different type of guider in the mutation stage to the search process.]{
\includegraphics[width=.3\textwidth]{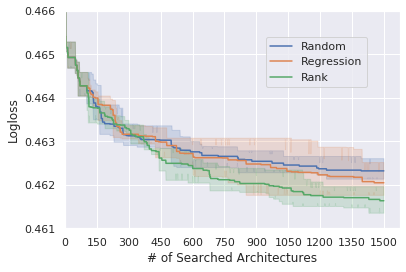}
    \label{fig:stage2}
}
\hspace{5pt}
\subfigure[Effects of different survivor selection objectives.]{
\includegraphics[width=.3\textwidth]{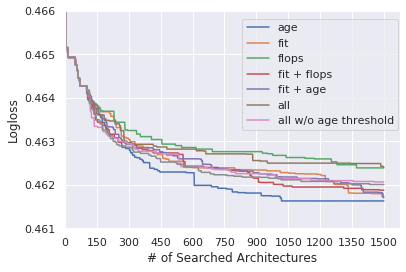}
    \label{fig:stage3}
}
\vspace{-0.5cm}
\caption{Analysis of the key hyperparameters in the three stages of AutoCTR} 
\vspace{-0.15cm}
\label{fig:stage} 
\end{figure*}


\subsubsection{Effects of Selection Intensity ($\lambda$) in Parent Selection.}  

We first analyze the influence of the selection intensity hyperparameter $\lambda$ in the parent selection stage by fixing the population size as 100. As discussed in section~\ref{sec:parent}, the higher the $\lambda$ is, the more intense the selection would become. We choose $\lambda=1, 5, 10, 25, 50$ and depict the curve of search effectiveness in Figure~\ref{fig:stage1}. With the increase of $\lambda$, the exploitation ability of AutoCTR generally increases while the exploration power decreases. It increases the initial search speed but would degrade the exploration ability in the long term. Under the current experimental setting, $\lambda=25$ seems to be a more balanced option between exploitation and exploration.

\subsubsection{Effects of Different Mutation Guider.} 

Secondly, we focus on the mutation stage and analyze the influence of the different types of guiders to the search efficacy. Three types of guiders are compared here: (1) random guider (random): it generates the candidate offspring by randomly selecting and mutating an operation of the selected parent architecture. (2) fitness-based guider (regression): it uses a gradient-boosted tree to conduct regression on the explored architectures and their performance. It selects the best architecture among the 100 randomly generated neighbors of the parent architecture as the new candidate offspring.  (3) rank-based guider (rank): the one we used in AutoCTR.  Different from the fitness-based guider, it learns the pairwise ranking relationship among the architectures rather than directly fit their performance. We fine-tune the tree-based learner for both fitness-based guider and rank-based guider, respectively. From Figure~\ref{fig:stage2}, we can observe that the AutoCTR with rank-based guider outperforms the other two. Although the fitness-based guider could improve search effectiveness comparing to the random strategy, it also suffers more on the overfitting issue, which results in the search variance to be large and makes it difficult to be tuned.

\begin{table}[t]\small
\vspace{-0.1cm}
\caption{Performance and complexity comparison of architectures found with different survivor selection objectives}
\vspace{-0.3cm}
\label{tab:stage3}
\begin{center}
\begin{tabular}{c|cc|cc}

\toprule
\multirow{2}{*}{Objective} & \multicolumn{2}{c|}{Performance} &  \multicolumn{2}{c}{Model Size (Million)}  \\
 & Logloss & AUC & \# Params & Flops \\\hline
$a_A$        &   0.4418    &   0.8010    &    21.97    &   16.62  \\
$r_A$        &   0.4417    &   0.8100    &    20.58    &   15.07   \\\hline  
$a_A + r_A$  &   0.4415    &   0.8103    &    19.77   &   11.85   \\
$a_A + c_A$  &   0.4418    &   0.8099    &    18.08    &   5.06    \\
$r_A + c_A$  &   0.4417    &   0.8101    &    19.59    &   11.10   \\\hline
$a_A + r_A + c_A$   &   0.4415  &   0.8103    &    20.50  & 14.73 \\
$a_A + r_A + c_A$ w/o threshold  &   0.4416    &   0.8102    &      19.35   &  10.14  \\\bottomrule
\end{tabular}
\end{center}
\vspace{-0.5cm}
\end{table}

\subsubsection{Effects of Different Survivor Selection Objectives.}

Finally, we explore the effect of adopting different objectives in the survivor selection stage. Figure~\ref{fig:stage3} and Table~\ref{tab:stage3} compare the search and final evaluation performance of AutoCTR with different search objectives. Except for the last objective, each of them adopts the age threshold  $\mathds{1}_{[a_A\le q]}$ described in Equation~\eqref{equ:multiobj}. We set the trade-off weights for each term as $0.5$. The results show that the age-based objective benefits more to the search speed comparing to the fitness-based objective. By adding the complexity constraint in the objective, the size of the best model explored could be reduced while the performance remains comparable.

\subsection{Discussion}

In this section, we visualize the best-explored architectures and analyze the importance of the block components learned from the tree-based guider. Some limitations and conjectures of the current study are discussed afterward to promote future exploration.

\subsubsection{Case Study.} 

\begin{figure}[t]
\centering
\includegraphics[height=4.3cm]{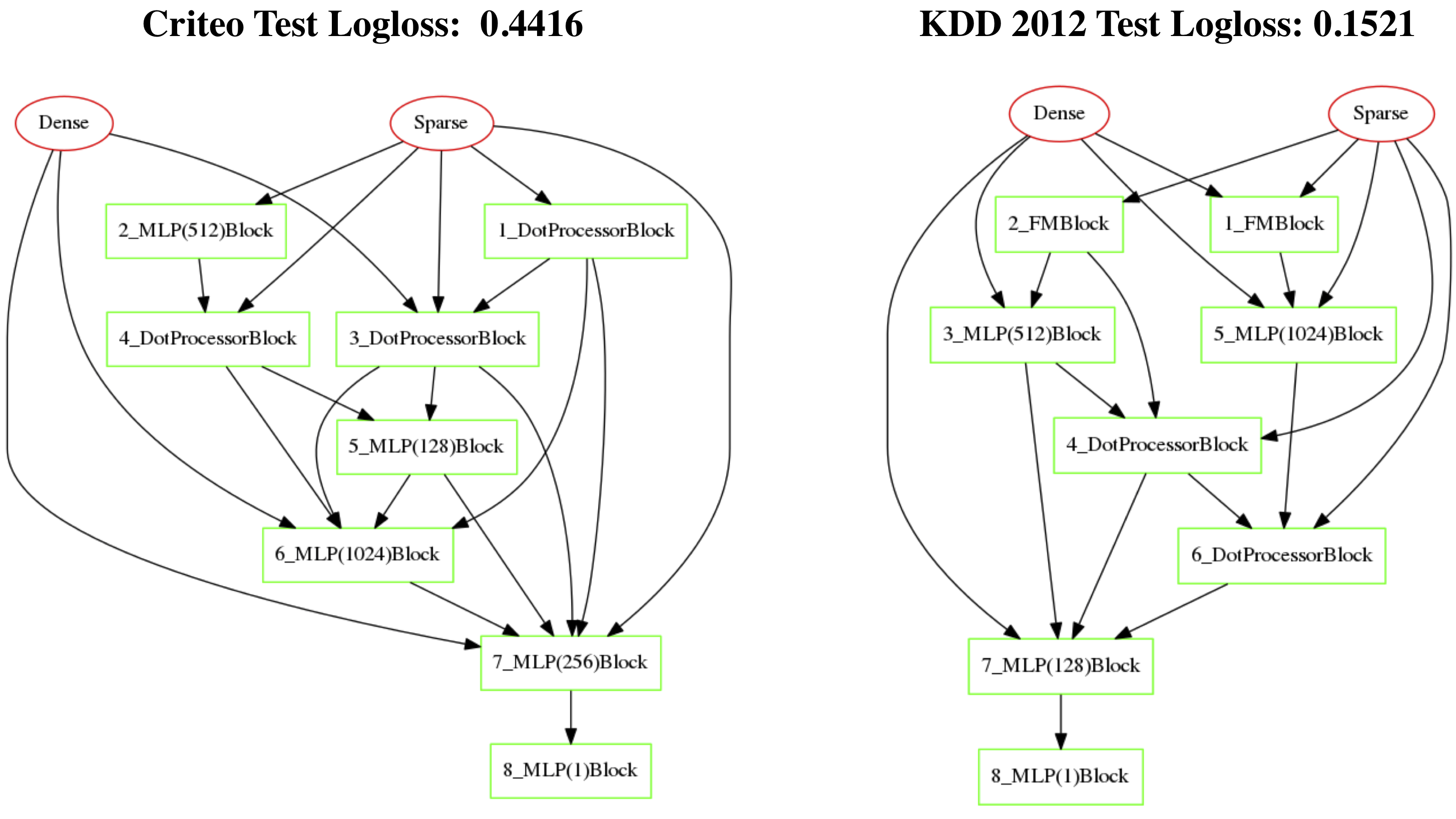} 
\vspace{-0.4cm}
\caption{Two architectures found by AutoCTR.}
\label{fig:case}
\vspace{-0.4cm}
\end{figure}

We first visualize two of the best architectures found by AutoCTR on Criteo and KDD datasets, respectively, in Figure~\ref{fig:case}. It shows that both of the architectures ensemble multiple FM and DP blocks and tend to adopt MLP blocks in the later stage. Besides, dense features prefer MLP block while sparse features prefer DP and FM in the early stages, which shows the ability of DP and FM in modeling sparse features explicitly. Moreover, the MLP blocks display a diamond structure, i.e., MLP layers in the middle of the graph are wider than the ones in both ends, which aligns with some analysis in existing works~\cite{zhang2016deep}: diamond networks are preferable to the increasing/decreasing width networks (triangular networks) and the constant width networks (rectangular networks).

\subsubsection{Interpretation of Important Blocks.}

\begin{figure}[t]
\centering
\vspace{-0.2cm}
\includegraphics[width=3in]{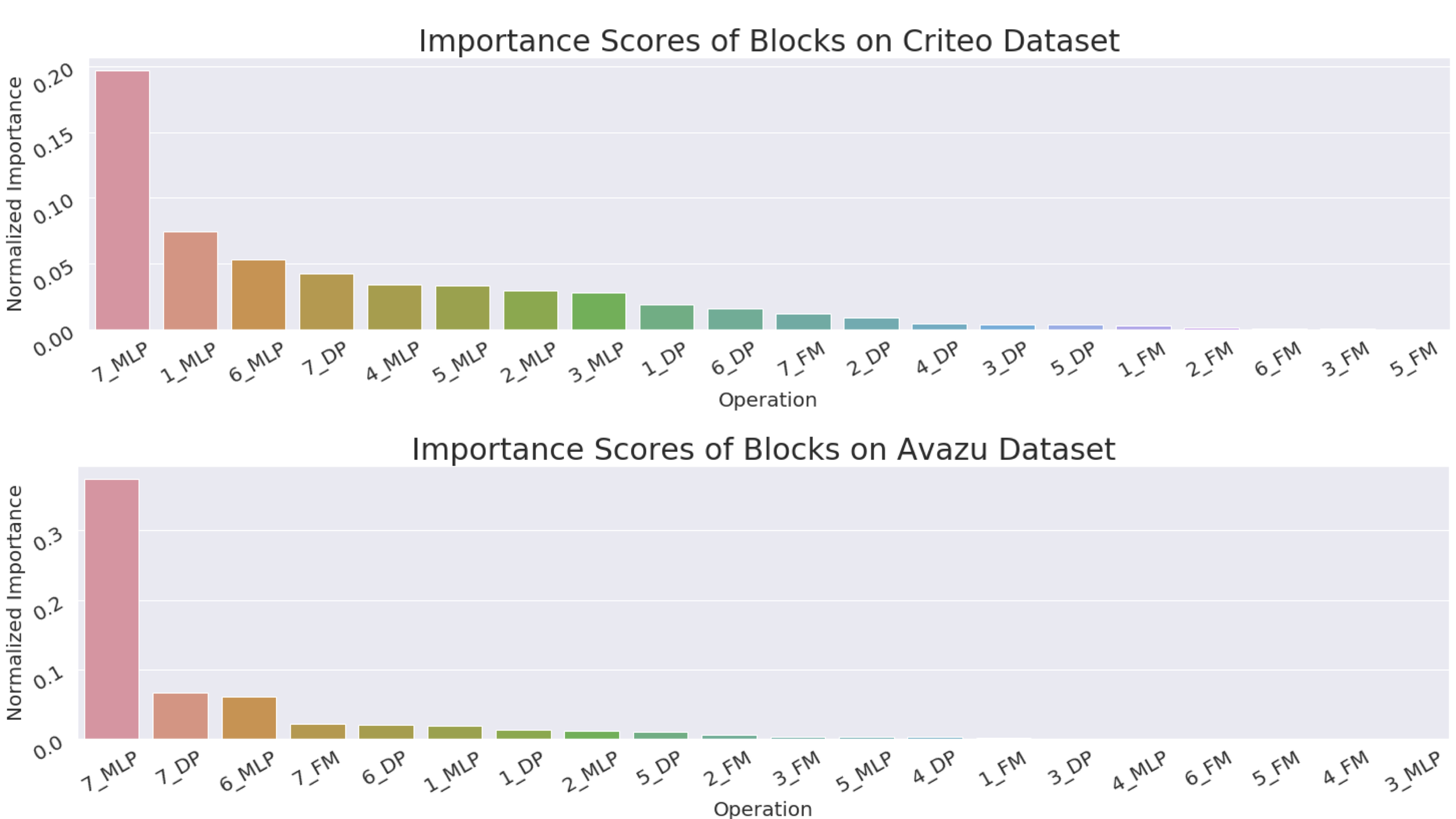} 
\vspace{-0.36cm}
\caption{Normalized importance scores of top-20 block type operations learned by AutoCTR guider on Criteo and Avazu} 
\label{fig:interpret}
\vspace{-0.5cm}
\end{figure}

To provide a better understanding of the block-type influence to the architectures, we display the feature importance of high influential components learned from the <architecture, performance rank> pairs on Criteo and Avazu via the tree-based guiders. We randomly select $10,000$ architectures with seven valid blocks in each of them to train the tree-based guider and display the importance score of the top-20 influential block types in Figure~\ref{fig:interpret}. The ``id\_type" tick below the x-axis indicates the topological order of a block and its type. We observe that: (1) the structure of the polar blocks based on the topological order have larger impacts compared with the middle ones; (2) MLP block dominates the architectures; (3) DP and FM blocks are relatively more impactful on Avazu than Criteo since Avazu only contains sparse features. We need to emphasize that this interpretation may be biased by 
the search-space design and the way of representing the architectures. Interpreting the NAS process and involving the interpretations into the architecture design could be promising.

\subsubsection{Limitations.} Despite the analysis discussed above, several limitations are mentioned here for future investigation. 

\noindent\textbf{Search Space.} Though the number of architectures contained in the search space is quite large ($> 10^{11}$), the number of block types we have currently explored is still limited. This also explains why Random and RL searchers could achieve acceptable performance. Moreover, the flexibility can be further enlarged via independent feature selection towards more dedicated and delicate interactions.

\noindent\textbf{Overfitting.} 
The overfitting issue is enlarged in the low-fidelity setting due to the limited subsample size and the stop criteria we adopted, i.e., search 1500 architectures for every searcher in each experiment. Although AutoCTR and LaNAS+ still show their superiority, the improvements compared with other searchers are weakened. One possible way to migrate this issue is to adopt early-stopping strategies or add regularizations for the search process~\cite{jiang2019neural, zela2020understanding}.

\section{Conclusions and Future Work}\label{con}
In this paper, we conduct a pilot study of automatically designing architectures for the CTR prediction task. We construct a hierarchical search space via wiring representative blocks extracting from human-crafted networks and explore the rank consistency among the architectures under the low-fidelity setting. A tailored evolutionary search algorithm with a multi-objective survivor selection strategy is proposed guided by an architectural-level learning-to-rank method. Experimental results on three benchmark datasets demonstrate the effectiveness of the proposed search algorithm and the feasibility of adopting low-fidelity estimation during the search phase. Future work includes considering independent feature interaction design, incorporating more diversified blocks to enrich the search space, as well as exploring more efficient search strategies.

\section{Acknowledgments}
\label{ack}
We would like to sincerely thank everyone who has provided their generous feedback for this work. Thank all the members of the Facebook personalization team for your feedback on the paper content, and thank the anonymous reviewers for their thorough comments and suggestions. This work is, in part, supported by DARPA Awards \#W911NF-16-1-0565 and \#FA8750-17-2-0116, and NSF Awards \#IIS-1657196 and \#IIS-1718840, granted to the co-author Xia Hu in his academic role at the Texas A\&M University. The views and conclusions contained in this paper are those of the authors and should not be interpreted as representing any funding agencies.

\bibliographystyle{ACM-Reference-Format}
\bibliography{autoctr}

\appendix

\section*{Reproducibility Supplementary Doc}

\section{Detailed Construction of the Three Example Building Blocks}\label{sec:block}

We elaborate on the three building blocks used in our final experiments here, i.e., dense MLP block, Dotprocessor Block, and FM Block. Other blocks, such as outer product block~\cite{wang2017deep} and self-attention block~\cite{song2019autoint},, etc., could also be easily integrated.

\noindent{\bf Dense MLP Block (MLP)} The MLP block serves as the most commonly used building block in literature. We design the dense MLP block similar to the one applied in the DLRM model~\cite{naumov2019deep}. The input of it will be concatenated into a single long vector and transformed via a multi-layer perceptron with the ReLU activation function. We set the layer to be $1$ for each MLP block to maximize the flexibility of the final constructed architectures. The width of each MLP block is searched within the unit set: \{32, 64, 128, 256, 512, 1024\}.

\noindent{\bf FM Block (FM)} The FM block refers to the factorization machine block~\cite{rendle2010factorization}. A conventional factorization machine model takes real-valued feature vector ${\bf x}$ as inputs, and outputs a single value via low-dimensional embedding and summation operations, as shown in Equation~\eqref{equ:fm}, where ${\bf v}$ denotes the embedding matrix, ${\bf w}$ is the transformation parameter. In the designed FM block, we assume all the input features are already transformed into the low-dimensional space and conduct the dot product  and summation operations directly. Three specific designs are listed here for dimensionality alignment, and search space robustness: (1) We concatenate all the dense input features, including dense raw features and the dense outputs from other blocks, into a single vector. (2) If the dimensions of sparse inputs and the concatenated dense input are conflicted, we will linearly embed them into the same size in advance. (3) If only dense features are obtained as inputs, the block will degenerate to a linear embedding block with a sum pooling afterward.
\begin{equation}\label{equ:fm}
\hat{y}({\bf x}) := w_0 + \sum_{i=1}^n w_ix_i + \sum_{i=1}^n \sum_{j=i+1}^n <{\bf v}_i, {\bf v}_j> x_i x_j.
\end{equation}

\noindent{\bf Dotprocessor Block (DP)} The DP block calculates the dot product of every pair of the input embeddings and concatenates the results into a single long vector. The self-interaction is also added, i.e., the dot product of each embedding vector and itself, so that when only dense features are collected as the block input, it is also feasible to proceed. In this case, this block degenerates into an element-wise square operator. Similar concatenation and dimensionality alignment strategy are used as for the FM block.

\section{Baseline Construction Details}\label{sec:baseline}

\subsection{Searcher Baselines} To prove the effectiveness of the proposed search algorithm, we select three representative NAS algorithms in existing work. We tailor random search and reinforcement learning based search to our designed search space and include a variation of LaNAS based on the proposed method and the parallel Monte Carlo Tree Search~\cite{chaslot2008parallel}.

\begin{itemize}[leftmargin=*]
\item {\bf Random Search.} Recent work in NAS has pointed out that random search could be a strong baseline, even comparing to the most advanced search algorithm~\cite{li2019random,sciuto2019evaluating}. We implement a random searcher as follows: given a random seed, in every search epoch, it randomly selects the four components described in section~\ref{sec:space} for each block and builds an architecture based on the topological order.

\item {\bf Reinforcement Learning Based Search (RL) }.  We adopt a single-layer RNN controller with length seven to generate architectures with seven intermediate blocks, as shown in Figure~\ref{fig:rl}. The output of each step decides the structure of the next block and serves as the recurrent input for the next step. Initial inputs and hidden state values are set as zero by default. We use the REINFORCE algorithm to update the controller and use the logloss difference between the current architecture and the best architecture so far on the validation set as the reward. The average reward is used as a baseline to reduce variance. An entropy term is added in the loss function to enhance the exploration diversity.

\begin{figure}[h]
\centering
\includegraphics[width=3.3in]{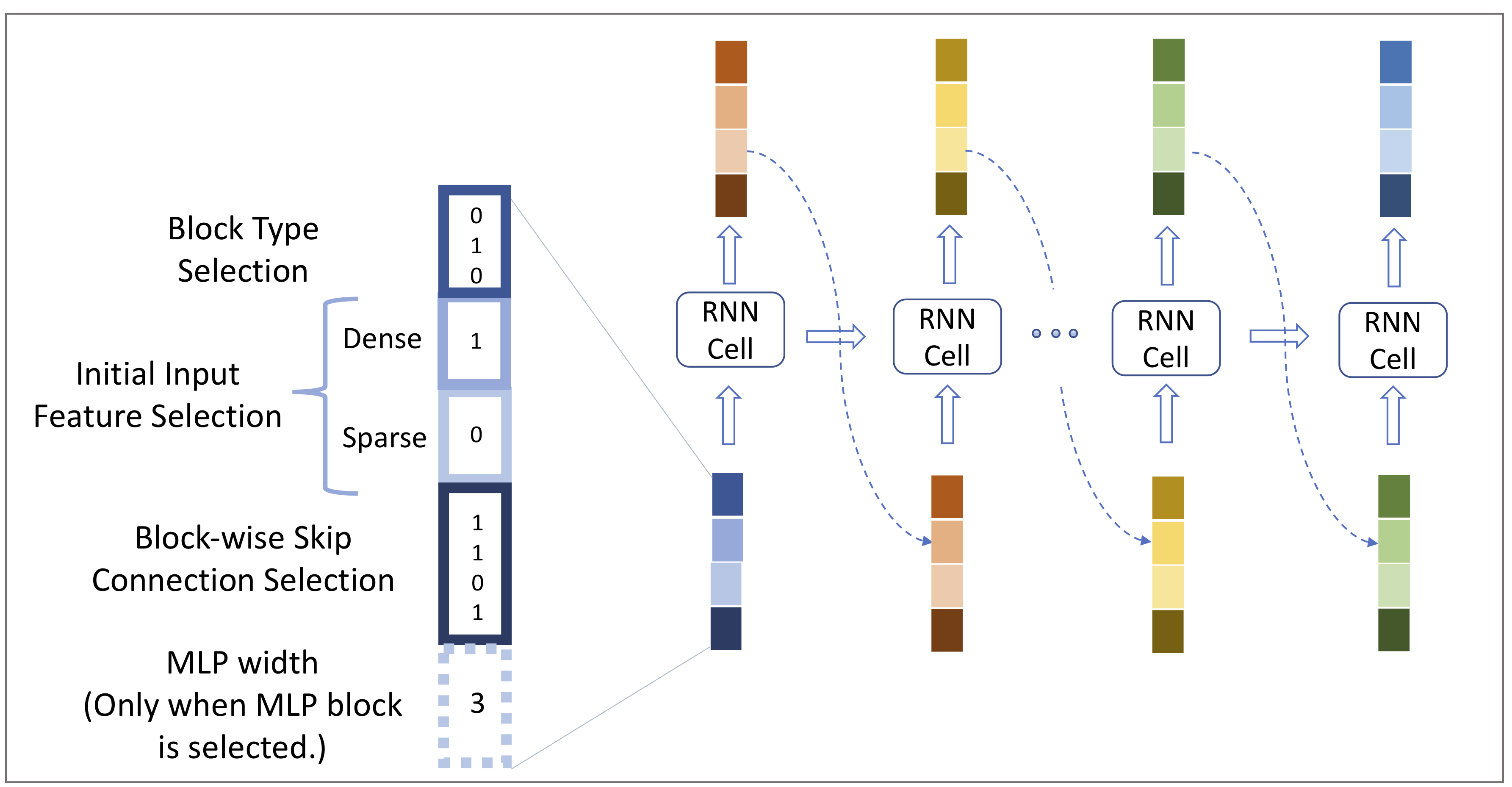} 
\caption{The illustration of the RL searcher}
\label{fig:rl}
\end{figure}

\item {\bf Latent Action Neural Architecture Search (LaNAS)}~\cite{wang2019sample}. A sample-efficient Monte Carlo Tree Search (MCTS) algorithm, which has been proven to be effective than various advanced search algorithms, including the Regularized evolutionary algorithm~\cite{real2019regularized} and Bayesian optimization algorithm in the image classification setting. The search strategy follows the standard MCTS algorithm with the upper-confidence bound (UCB) policy. The search space is partitioned via customizable regressors contained in the nodes of the tree. The search algorithm could narrow down to a specific subspace when traversing from the root to the leaf. Each regressor is updated based on the architectures in the corresponding node and their validation logloss.

\item {\bf LaNAS+}. We improve LaNAS from two perspectives. 

\begin{enumerate}
    \item Convert its rollout sampling policy from random sampling to rank-based method borrowing the idea from the evolutionary method. Random rollout policy could be slow due to the lack of exploitation. Following the original paper, the random rollout policy here means when we have selected a path in the tree and collected the constraints in each node along the path to narrow down to a specific subspace, the way to sample an architecture from this subspace is naively random sampling. However, this could potentially cause a high rejection rate due to the complicacy of sampling from a non-convex polytope and may lose the usage of existed good architectures. To address the problem, we design an evolutionary style rollout policy: every time after narrowing down to a specific subspace, we retrieve the currently searched architectures within this subspace and set them as the population. Then we do the same thing as we do in the AutoCTR, i.e., select a subset of candidates, pick the best one, and mutate it to a new architecture. 
    \item Introduce virtual loss to enable parallel training of multiple architectures to improve the search speed. Within each update internal of the LaNAS searcher, we are supposed to sample multiple architectures in order to explore different tree branches. After each architecture is sampled, the back-propagate step requires the logloss of it to update the tree statistics so that it could conduct exploration on different branches rather than always exploring one single path. However, this may prevent the searching process from being parallelly proceeded since each new architecture can only be sampled after the evaluation of the former one. To circumvent this problem, a virtual loss~\cite{chaslot2008parallel} could be applied as padding during the back-propagate step to ensure the feasible adoption of parallel training and evaluation of multiple architectures. The virtual losses will be removed after the true loss is obtained.

\end{enumerate}

\end{itemize}

\subsection{SOTA Human-Crafted Networks} We select three representative human-designed networks to examine if the explored network is able to achieve comparable accuracy or even beat the SOTA human-crafted networks.
\begin{itemize}[leftmargin=*]

\item {\bf DeepFM}~\cite{guo2017deepfm}: a two-tower model composed of a six-layer MLP with 1024 hidden units in each layer and factorization machine block, the sparse input embeddings are shared between the two components, and the output of the two components are linearly embedded with sigmoid transformation as the final prediction.

\item {\bf DLRM}~\cite{naumov2019deep}: a SOTA MLP-based recommendation model. It encodes dense features and sparse look-up embeddings with two MLP modules and embeds their outputs with a top-level MLP module jointly. We adopt two single-layer MLP on the dense and sparse features respectively and stack a six-layer MLP on top of them. All MLP layers are with 1024 units except the final one.

\item {\bf AutoInt+}~\cite{song2019autoint}: a two-tower model composed of a four-layer MLP with 1024 hidden units in each layer and three-layer self-interaction layer, which adopts the multi-head self-attention schema to learn high-order feature interactions. The outputs of the two components are linearly embedded with sigmoid transformation into the final prediction. 
\end{itemize}

\section{Experimental Settings}\label{sec:expset}

\subsection{The Statistics of Three Benchmarks}\label{sec:data}

\begin{table}[H]\small
\caption{Statistics of Datasets}
\label{tab:dataset}
\begin{center}
\begin{tabular}{c|c|c|c|c}
\toprule
 Datasets   & \# Samples &  \begin{tabular}{@{}c@{}} \# Dense \\ Features\end{tabular} &  \# \begin{tabular}{@{}c@{}} \# Sparse \\ Features\end{tabular} & \begin{tabular}{@{}c@{}} Total Cardinality of  \\ Sparse Features\end{tabular} 
\\ \hline
Criteo\footref{criteo}        &  45,840,617   & 13  & 26   &  998,960      \\ \hline
Avazu\footref{avazu}        &  40,428,967   &  0   & 23  &  1,544,488    \\ \hline
KDD\footref{kdd}    &  149,639,105 &  3   & 10  &  6,019,086     \\ 
\bottomrule
\end{tabular}
\end{center}
\end{table}

We would like to gratefully acknowledge the organizers of KDD Cup 2012 track 2 as well as the contributors of Criteo and Avazu for making these CTR prediction benchmarks publicly available.

\subsection{Hyperparameter Settings}

 We elaborate on the detailed hyperparameter settings adopted in the experiments in this section. The three random seeds used in the experiments are $42$, $2019$, and $1234$, respectively.

\begin{itemize}[leftmargin=*]
 \item \textbf{Search Phase Training:}
\textit{Adam} and \textit{Sparse Adam} optimizers are adopted for dense and sparse features, respectively, in training. The batch size is set as 4096. The learning rate is $0.001$. The hash size is $10^4$ for all sparse features. The embedding tables for sparse features are randomly initialized based on a normal distribution with $0$ mean and $0.01$ standard deviation. The embedding size for each sparse feature is $16$.

 \item \textbf{Final-Fit Phase Training:} Except for the hash sizes of all sparse features that are set back to their original cardinality, and the dataset is the full data, all the other settings are the same with the ones adopted in the search phase. 
\end{itemize}

\noindent Hyperparameter settings for the searchers are elaborated as follows.

\begin{itemize}[leftmargin=*]
    \item \textbf{RL searcher:} the input encoding size and the LSTM hidden embeddings size are both set to 10, the trade-off hyperparameter of the entropy term is set to be 0.1. 
    
    \item \textbf{LaNAS:} the tree-depth is set to 5. In the search phase, the update step is done once after 20 architectures are evaluated. The UCB trade-off parameter is set as 0.5. The space split classifier is defined as ridge regression with 0.1 regularization hyperparameter. 
    
    \item \textbf{LaNAS+:} besides the hyperparameters  mentioned above in LaNAS, the candidate size for the modified rollout policy in LaNAS+ is set as to be half of the architectures contained in a selected region. The virtual loss is set to be the mean logloss of the architectures contained in each leaf node.  

    \item \textbf{AutoCTR:} the population size is set to be 100, and the survivor selection threshold $q$ is set as $200$. The trade-off hyperparameters $\mu_1, \mu_2, \mu_3$ are set to be 1, 0.1, and 0.1, respectively. The parent selection trade-off hyperparameter $\lambda$ is set to be $10$. The guider is implemented with the lightgbm package with NDCG@3 as the early-stop evaluation metric.  
 \item \textbf{AutoCTR (warm):} the warm-start embedding for each dataset is achieved from a four-layer MLP architecture with units: 128-1024-128-1, which is pretrained on each full dataset. 
\end{itemize}

\subsection{Software and Hardware Descriptions}\label{sec:hardware}
All the deep learning related frameworks are implemented with the PyTorch package\footnote{\url{https://pytorch.org}}. Every single search experiment is run on a single GPU (NVIDIA GeForce RTX 2080 Ti) with three architectures parallelly trained on it. Multi-core CPUs are used for data preprocessing and searcher training. Specifically, we use five cores for data preprocessing and the searcher training in the search phase, and 5 CPU cores + 1 GPU for the final fit of each discovered architecture after searching. We adopt parallel CPU-GPU training for searching, and the training speed is accelerated by loading, preprocessing, and saving the data batches in the GPU memory. The evolutionary guider is implemented with the lightgbm package\footnote{\url{https://lightgbm.readthedocs.io/}}, and the FLOPs are calculated based on the thop package\footnote{\url{https://github.com/Lyken17/pytorch-OpCounter}}. Plots in the case study are drawn with the graphviz package\footnote{\url{https://graphviz.readthedocs.io}}.

\end{document}